\documentclass[letterpaper,12pt,reqno]{article}
\usepackage{authblk}

\usepackage{authblk}
\usepackage{amsthm,amssymb,amsmath}
\usepackage{graphicx,hyperref}
\usepackage{eqnarray,array}
\usepackage{pdfsync}
\usepackage{tikz}
\usepackage{booktabs}
\usepackage{color,xcolor}
\usepackage{tikz,3dplot}
\usepackage{subfig,caption}
\usepackage{chngpage}
\usepackage[top=1.5in, bottom=1.5in, left=1.5in, right=1.5in]{geometry}
\usetikzlibrary{arrows,positioning}

\newtheorem{thm}{Theorem}

\theoremstyle{definition}

\newtheorem{lcon}{Linearity Condition}

\theoremstyle{remark}

\title{\Large \bf Partial Sliced Inverse Regression for Quality-Relevant Multivariate Statistical Process Monitoring}

\date{}

\author[1]{Yue Yu}
\author[2]{Zhijie Sun}
\affil[1]{Department of Mathematics, Statistics, and Computer Science, University of Illinois at Chicago}
\affil[2]{Mork Family Department of Chemical Engineering and Materials Science, University of Southern California}

\begin{document}

\maketitle

\begin{abstract}
This paper introduces a popular dimension reduction method, sliced inverse regression (SIR), into multivariate statistical process monitoring. Provides an extension of SIR for the single-index model by adopting the idea from partial least squares (PLS). Our partial sliced inverse regression (PSIR) method has the merit of incorporating information from both predictors ($\mathbf{x}$) and responses ($\mathbf{y}$), and it has capability of handling large, nonlinear, or ``$n<p$'' dataset. Two statistics with their corresponding distributions and control limits are given based on the X-space decomposition of PSIR for the purpose of fault detection in process monitoring. Simulations showed PSIR outperformed over PLS and SIR for both linear and nonlinear model.
\end{abstract}
%

\section{Introduction}

Quality-relevant multivariate statistical process monitoring has been studied for a long time, it is one of the most active research topics in the interdisciplinary area of statistical and engineering during the past two decades. And it has been widely used in chemical engineering, manufacture, healthcare, pharmaceutical, electronics, and agriculture. There are many literatures in this area, for example, Kresta et al. \cite{kresta1991multivariate}, Nomikos and MacGregor \cite{nomikos1995multivariate}, Dunia and Qin \cite{dunia1998subspace}, Qin \cite{qin2003statistical} and Li, et al. \cite{li2010geometric}.

The concept of the quality-relevant multivariate statistical process monitoring is to monitor the abnormal observations in the measurements, which is usually assumed to have a multivariate normal distribution. And the responses are the quality variables associated with the processing conditions. The quality variables are often assumed to be correlated with the measurements, but the relationship is unknown and they may also be affected by some other independent factors. Therefore, the better way to detect the fault in a process is to consider the information from both measurements and quality variables.

Two of the latest methods to analysis multivariate statistical process are principal component analysis (PCA) and partial least square (PLS). Dunia and Qin \cite{dunia1998subspace} and Qin \cite{qin2003statistical} gave a complete introduction for using the latent space decomposition via PCA to perform the fault detection in process monitoring. Li, et al. \cite{li2010geometric} extended such methodology to the latent space of PLS, which integrates the information from both the measurements and quality variables.

In 1991, Li \cite{li1991sliced} and Duan and Li \cite{duan1991slicing} proposed a new way of thinking in the regression analysis, called sliced inverse regression (SIR). SIR reverses the role of responses $\mathbf{x}$ and predictors $\mathbf{y}$. Traditional regression methods mainly studies the conditional density $f(\mathbf{y}|\mathbf{x})$. SIR gains the information of the variation of predictors $\mathbf{x}$ when responses $\mathbf{y}$ change, by studying the conditional density $h(\mathbf{x}|\mathbf{y})$. Usually the dimension of the responses is far more less than the dimension of the predictors, hence, it is a way to avoid the ``curse of dimensionality''.

The basic SIR algorithm considers the inverse conditional expectation $E(\mathbf{x}|\mathbf{y})$. There are several extensions of SIR using the high order moments of the inverse function. For example, SIR-II (Li, \cite{li1991rejoinder}), sliced average variance estimator (SAVE) (Cook and Weisberg, \cite{cook1991discussion}) and sliced inverse moment regression (SIMR) (Ye and Yang, \cite{ye2010sliced}). There are also some extensions of SIR for the highly collinearity data and $n<p$ problems, for example, regularized sliced inverse regression (Zhong, et al., \cite{Zhong2005RSIR}, Li and Ying, \cite{LiYin2008}) and partial inverse regression (Li, et al., \cite{li2007partial}).

The objective of this paper is to apply the partial sliced inverse regression in multivariate process monitoring. The remaining of the paper is organized as follows. Section 2 reviews the SIR and PSIR algorithms with their conditions and selection of parameters. Section 3 introduces the X-space decomposition for both PSIR and PLS. The process monitoring based on PSIR is presented in Section 4, two fault detection statistics and a combined index with their corresponding distributions and control limits are presented. Section 5 provides a simulation study to compare the performance of PSIR and PLS. The conclusion and discussion are given in the last section.

\section{Dimension Reduction in Regression}

\subsection{Sliced Inverse Regression (SIR)}

SIR is a model free dimension reduction method introduced by Li \cite{li1991sliced} and Duan and Li \cite{duan1991slicing}. Consider the single-response case first. The model takes the form of
\begin{equation}\label{e1}
y=g(\beta_1^T\mathbf{x},\beta_2^T\mathbf{x},\dots,\beta_K^T\mathbf{x},\epsilon),
\end{equation}
where $y$ is assumed to be univariate, $\mathbf{x}$ is a $p$-dimensional column vector, and the random error $\epsilon$ is unknown and independent of $\mathbf{x}$. $g(\cdot)$ is an arbitrary unknown function. Model \eqref{e1} means that $y$ depends on $\mathbf{x}$ only through the $K$-dimensional subspace spanned by projection vectors $\beta_1, \dots, \beta_K$, known as the effective dimension reducing directions (e.d.r.-directions).

Unlike traditional regression methods, SIR intends to collect the information on how $\mathbf{x}$ changes along with $y$. Hence, instead of estimating the forward regression function $\eta(\mathbf{x})=E(y|\mathbf{x})$, the inverse regression methodology is to consider $\xi(y)=E(\mathbf{x}|y)$. Compared to the forward regression function, the inverse regression function depends on one-dimension $y$, which makes the analysis much more easier.

Duan and Li \cite{duan1991slicing} showed that the e.d.r.-directions can be estimated by solving
\begin{equation}\label{e3}
\mathrm{Cov}\big(E(\mathbf{x}|y)\big)\beta_j=\lambda_j \mathrm{Cov}(\mathbf{x}) \beta_j,
\end{equation}
where $\lambda_j$ is the $j$-th eigenvalue and $\beta_j$ is the corresponding eigenvector with respect to $\mathrm{Cov}(\mathbf{x})$. The covariance matrices can be replaced by their sample counterparts during the forecasting procedure.

For the given data $(\mathbf{y}, \mathbf{x}_i)$, $i =1,\dots,p$, $\mathbf{x}_i,\ \mathbf{y} \in \mathbb{R}^{n\times 1}$,
the SIR algorithm can be described as following:

\begin{enumerate}

  \item Normalize predictors $\mathbf{x}_i$ to zero means and identity covariance:
  $$\mathbf{z}_i=\hat{\Sigma}_{\mathbf{x}}^{-1/2}(\mathbf{x}_i-\bar{\mathbf{x}}),$$
  where $\hat{\Sigma}_{\mathbf{x}}=\sum_{i=1}^p (\mathbf{x}_i-\bar{\mathbf{x}})(\mathbf{x}_i-\bar{\mathbf{x}})^T/p$, $\bar{\mathbf{x}}=\sum_{i=1}^p \mathbf{x}_i/p$;

  \item Sort the values of $\mathbf{y}$ and then partition them into $H$ slices;

  \item Distribute $\mathbf{z}_i$ into these $H$ slices and compute their covariance:
  \begin{equation*}
  \Sigma_{\rho}=\sum_{h=1}^H \hat{\rho}_h \bar{\mathbf{X}}_h\bar{\mathbf{X}}^T_h,
  \end{equation*}
  where $\hat{\rho}_h$ is the proportion of observations falling into slice $h$, and $\bar{\mathbf{X}}_h=\sum_{i=1}^p I_{\mathbf{z}_i \in h}\mathbf{z}_i/n_i$;

  \item Find the eigenvector of $\Sigma_{\rho}$, $\hat{\eta}_1, \hat{\eta}_2, \dots, \hat{\eta}_K$. The e.d.r.-directions are
  $$\hat{\beta}_k=\hat{\Sigma}_{\mathbf{x}}^{-1/2}\hat{\eta}_k,\quad k=1,2,\dots,K.$$

\end{enumerate}

A chi-square test was given by Li \cite{li1991sliced} to determine the number of significant e.d.r.-directions $K$. 

Li \cite{li1991sliced} proved the Fisher consistency of SIR by assuming the following linearity condition.

\begin{lcon}\label{cond1}
For any $b \in \mathbb{R}^p$, the conditional expectation
$E(b^T\mathbf{x}|\allowbreak \beta_1^T\mathbf{x},\dots,\beta_K^T\mathbf{x})$ is
linear in $\beta_1^T\mathbf{x},\dots,\beta_K^T\mathbf{x}$.
\end{lcon}

\begin{thm}[Li \cite{li1991sliced}]\label{thm1}
Assume Linearity Condition \ref{cond1}, the centered inverse regression curve $E(\mathbf{x}|y)$ is contained in the space spanned by $\Sigma_{\mathbf{x}}\boldsymbol{\beta}_j,\ j=1,\dots,K$, where $\Sigma_{\mathbf{x}}$ is the covariance matrix of $\mathbf{x}$.
\end{thm}

Although the Linearity Condition \ref{cond1} is not easy to be verified, it can be shown when $\mathbf{x}$ is elliptically symmetrically distributed, and particularly, when $\mathbf{x}$ follows a multivariate normal distribution, the linearity condition holds \cite{eaton1986characterization}.

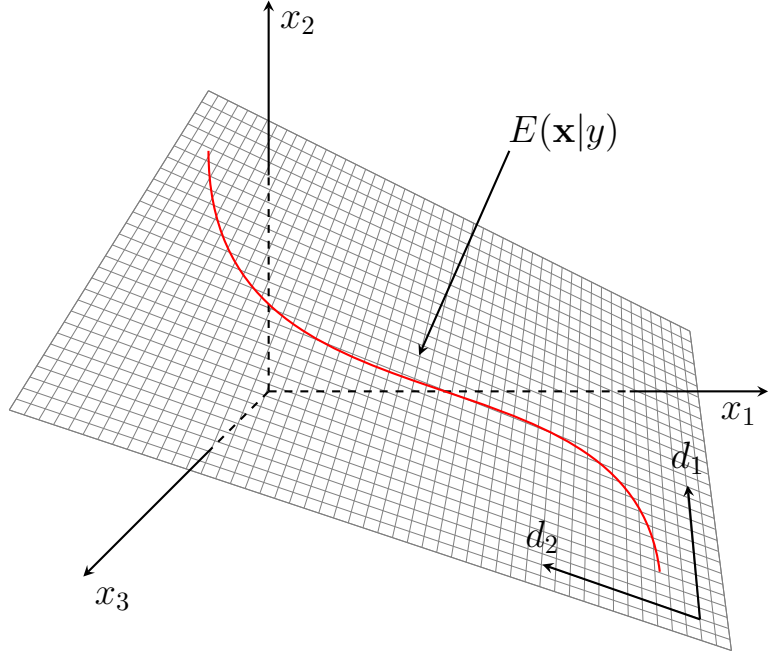
\begin{figure}[!tpb]
\begin{center}
\tdplotsetmaincoords{60}{110}
\pgfmathsetmacro{\rvec}{.8}
\pgfmathsetmacro{\thetavec}{30}
\pgfmathsetmacro{\phivec}{60}
\begin{tikzpicture}[scale=0.8]
\tikzstyle{every node}=[font=\large]
\coordinate (O) at (0,0,0);
\tdplotsetcoord{P}{\rvec}{\thetavec}{\phivec}
\draw[gray, thin,-] (-4,5,0) -- (4,1,0);
\draw[gray, thin, -] (-5,2,6) -- (7,-2,6);
\draw[gray, thin, -] (-4,5,0) -- (-5,2,6);
\draw[gray, thin, -] (7,-2,6) -- (4,1,0);
\draw[gray, very thin, -] (-4,5,0) -- (-5,2,6);
\draw[gray, very thin, -] (-3.8367,4.9184,0) -- (-4.7551,1.9184,6);
\draw[gray, very thin, -] (-3.6735,4.8367,0) -- (-4.5102,1.8367,6);
\draw[gray, very thin, -] (-3.5102,4.7551,0) -- (-4.2653,1.7551,6);
\draw[gray, very thin, -] (-3.3469,4.6735,0) -- (-4.0204,1.6735,6);
\draw[gray, very thin, -] (-3.1837,4.5918,0) -- (-3.7755,1.5918,6);
\draw[gray, very thin, -] (-3.0204,4.5102,0) -- (-3.5306,1.5102,6);
\draw[gray, very thin, -] (-2.8571,4.4286,0) -- (-3.2857,1.4286,6);
\draw[gray, very thin, -] (-2.6939,4.3469,0) -- (-3.0408,1.3469,6);
\draw[gray, very thin, -] (-2.5306,4.2653,0) -- (-2.7959,1.2653,6);
\draw[gray, very thin, -] (-2.3673,4.1837,0) -- (-2.551,1.1837,6);
\draw[gray, very thin, -] (-2.2041,4.102,0) -- (-2.3061,1.102,6);
\draw[gray, very thin, -] (-2.0408,4.0204,0) -- (-2.0612,1.0204,6);
\draw[gray, very thin, -] (-1.8776,3.9388,0) -- (-1.8163,0.9388,6);
\draw[gray, very thin, -] (-1.7143,3.8571,0) -- (-1.5714,0.8571,6);
\draw[gray, very thin, -] (-1.551,3.7755,0) -- (-1.3265,0.7755,6);
\draw[gray, very thin, -] (-1.3878,3.6939,0) -- (-1.0816,0.6939,6);
\draw[gray, very thin, -] (-1.2245,3.6122,0) -- (-0.8367,0.6122,6);
\draw[gray, very thin, -] (-1.0612,3.5306,0) -- (-0.5918,0.5306,6);
\draw[gray, very thin, -] (-0.898,3.449,0) -- (-0.3469,0.449,6);
\draw[gray, very thin, -] (-0.7347,3.3673,0) -- (-0.102,0.3673,6);
\draw[gray, very thin, -] (-0.5714,3.2857,0) -- (0.1429,0.2857,6);
\draw[gray, very thin, -] (-0.4082,3.2041,0) -- (0.3878,0.2041,6);
\draw[gray, very thin, -] (-0.2449,3.1224,0) -- (0.6327,0.1224,6);
\draw[gray, very thin, -] (-0.0816,3.0408,0) -- (0.8776,0.0408,6);
\draw[gray, very thin, -] (0.0816,2.9592,0) -- (1.1224,-0.0408,6);
\draw[gray, very thin, -] (0.2449,2.8776,0) -- (1.3673,-0.1224,6);
\draw[gray, very thin, -] (0.4082,2.7959,0) -- (1.6122,-0.2041,6);
\draw[gray, very thin, -] (0.5714,2.7143,0) -- (1.8571,-0.2857,6);
\draw[gray, very thin, -] (0.7347,2.6327,0) -- (2.102,-0.3673,6);
\draw[gray, very thin, -] (0.898,2.551,0) -- (2.3469,-0.449,6);
\draw[gray, very thin, -] (1.0612,2.4694,0) -- (2.5918,-0.5306,6);
\draw[gray, very thin, -] (1.2245,2.3878,0) -- (2.8367,-0.6122,6);
\draw[gray, very thin, -] (1.3878,2.3061,0) -- (3.0816,-0.6939,6);
\draw[gray, very thin, -] (1.551,2.2245,0) -- (3.3265,-0.7755,6);
\draw[gray, very thin, -] (1.7143,2.1429,0) -- (3.5714,-0.8571,6);
\draw[gray, very thin, -] (1.8776,2.0612,0) -- (3.8163,-0.9388,6);
\draw[gray, very thin, -] (2.0408,1.9796,0) -- (4.0612,-1.0204,6);
\draw[gray, very thin, -] (2.2041,1.898,0) -- (4.3061,-1.102,6);
\draw[gray, very thin, -] (2.3673,1.8163,0) -- (4.551,-1.1837,6);
\draw[gray, very thin, -] (2.5306,1.7347,0) -- (4.7959,-1.2653,6);
\draw[gray, very thin, -] (2.6939,1.6531,0) -- (5.0408,-1.3469,6);
\draw[gray, very thin, -] (2.8571,1.5714,0) -- (5.2857,-1.4286,6);
\draw[gray, very thin, -] (3.0204,1.4898,0) -- (5.5306,-1.5102,6);
\draw[gray, very thin, -] (3.1837,1.4082,0) -- (5.7755,-1.5918,6);
\draw[gray, very thin, -] (3.3469,1.3265,0) -- (6.0204,-1.6735,6);
\draw[gray, very thin, -] (3.5102,1.2449,0) -- (6.2653,-1.7551,6);
\draw[gray, very thin, -] (3.6735,1.1633,0) -- (6.5102,-1.8367,6);
\draw[gray, very thin, -] (3.8367,1.0816,0) -- (6.7551,-1.9184,6);
\draw[gray, very thin, -] (4,1,0) -- (7,-2,6);

\draw[gray, very thin, -] (-4,5,0) -- (4,1,0);
\draw[gray, very thin, -] (-4.0345,4.8966,0.2069) -- (4.1034,0.8966,0.2069);
\draw[gray, very thin, -] (-4.069,4.7931,0.4138) -- (4.2069,0.7931,0.4138);
\draw[gray, very thin, -] (-4.1034,4.6897,0.6207) -- (4.3103,0.6897,0.6207);
\draw[gray, very thin, -] (-4.1379,4.5862,0.8276) -- (4.4138,0.5862,0.8276);
\draw[gray, very thin, -] (-4.1724,4.4828,1.0345) -- (4.5172,0.4828,1.0345);
\draw[gray, very thin, -] (-4.2069,4.3793,1.2414) -- (4.6207,0.3793,1.2414);
\draw[gray, very thin, -] (-4.2414,4.2759,1.4483) -- (4.7241,0.2759,1.4483);
\draw[gray, very thin, -] (-4.2759,4.1724,1.6552) -- (4.8276,0.1724,1.6552);
\draw[gray, very thin, -] (-4.3103,4.069,1.8621) -- (4.931,0.069,1.8621);
\draw[gray, very thin, -] (-4.3448,3.9655,2.069) -- (5.0345,-0.0345,2.069);
\draw[gray, very thin, -] (-4.3793,3.8621,2.2759) -- (5.1379,-0.1379,2.2759);
\draw[gray, very thin, -] (-4.4138,3.7586,2.4828) -- (5.2414,-0.2414,2.4828);
\draw[gray, very thin, -] (-4.4483,3.6552,2.6897) -- (5.3448,-0.3448,2.6897);
\draw[gray, very thin, -] (-4.4828,3.5517,2.8966) -- (5.4483,-0.4483,2.8966);
\draw[gray, very thin, -] (-4.5172,3.4483,3.1034) -- (5.5517,-0.5517,3.1034);
\draw[gray, very thin, -] (-4.5517,3.3448,3.3103) -- (5.6552,-0.6552,3.3103);
\draw[gray, very thin, -] (-4.5862,3.2414,3.5172) -- (5.7586,-0.7586,3.5172);
\draw[gray, very thin, -] (-4.6207,3.1379,3.7241) -- (5.8621,-0.8621,3.7241);
\draw[gray, very thin, -] (-4.6552,3.0345,3.931) -- (5.9655,-0.9655,3.931);
\draw[gray, very thin, -] (-4.6897,2.931,4.1379) -- (6.069,-1.069,4.1379);
\draw[gray, very thin, -] (-4.7241,2.8276,4.3448) -- (6.1724,-1.1724,4.3448);
\draw[gray, very thin, -] (-4.7586,2.7241,4.5517) -- (6.2759,-1.2759,4.5517);
\draw[gray, very thin, -] (-4.7931,2.6207,4.7586) -- (6.3793,-1.3793,4.7586);
\draw[gray, very thin, -] (-4.8276,2.5172,4.9655) -- (6.4828,-1.4828,4.9655);
\draw[gray, very thin, -] (-4.8621,2.4138,5.1724) -- (6.5862,-1.5862,5.1724);
\draw[gray, very thin, -] (-4.8966,2.3103,5.3793) -- (6.6897,-1.6897,5.3793);
\draw[gray, very thin, -] (-4.931,2.2069,5.5862) -- (6.7931,-1.7931,5.5862);
\draw[gray, very thin, -] (-4.9655,2.1034,5.7931) -- (6.8966,-1.8966,5.7931);
\draw[gray, very thin, -] (-5,2,6) -- (7,-2,6);

\draw[dashed,thick,-] (-3,0,0) -- (3,0,0);
\draw[thick,-stealth] (3,0,0) -- (5.3,0,0) node[anchor=north east]{$x_1$};
\draw[dashed,thick,-] (-3,0,0) -- (-3,3.6,0);
\draw[thick,-stealth] (-3,3.6,0) -- (-3,6.5,0) node[anchor=north west]{$x_2$};
\draw[dashed,thick,-] (-3,0,0) -- (-3,0,2.42);
\draw[thick,-stealth] (-3,0,2.42) -- (-3,0,8) node[anchor=north west]{$x_3$};

\draw[thick,-stealth] (6.17,-1.79,5.20) -- (5.08,-0.45,2.90) node[anchor=south]{$d_1$};
\draw[thick,-stealth] (6.17,-1.79,5.20) -- (1.55,-2.88,0) node[anchor=south]{$d_2$};

\draw [thick,red] (-4,4,0) .. controls (-4,-1,0) and (3,1,0) .. (3.5,-3,0);
\draw[thick,-stealth] (1,4,0) -- (-0.5,0.6,0);
\node[right] at (0.8,4.3,0) {$E(\mathbf{x}|y)$};
\end{tikzpicture}
\caption{Inverse Regression Curve in a Three-Dimensional Space}\label{f0.5}
\end{center}
\end{figure}

Figure \ref{f0.5} shows a three-dimensional case when $\mathbf{x}=\big(x_1,x_2,x_3\big)'$, since the inverse regression function $E(\mathbf{x}|y)$ is a function of $y$, it draws a curve in the three-dimensional space when $y$ changes. Theorem \ref{thm1} indicates that such curve is located exactly on the plane spanned by two directions $d_1$ and $d_2$ from $\Sigma_{\mathbf{x}}\boldsymbol{\beta}_j,\ j=1,2$,  assuming $K=2$.

The methodology of sliced inverse regression can be easily extended to the multiple responses case. For example, if $\mathbf{y}$ has $m$ variables, each variable can be partitioned into $H_i,\ i=1,\dots,m$ slices, yielding a total of $H=H_1\times H_2\times \dots \times H_m$ slices. Then the slice means and covariance of slice means can be calculated similarly. Note that the total number of slices will inflate quickly when the dimension of $\mathbf{y}$ increases. Li, et al. \cite{li2003dimension} discussed such problem throughly, and they found a way to reduce the dimension for $\mathbf{y}$ using the same approach as SIR. One may refer to their article if interested.


\subsection{Partial Sliced Inverse Regression (PSIR)}
To begin with, the single-index model is considered, and the multiple-index model will be discussed later. The single-index model includes only one linear combination of the predictors $\mathbf{x}$. It is adequate enough for most of the cases in industry if the dimension of the predictors is not large.
\begin{equation}\label{e2}
y=g(\beta^T\mathbf{x},\epsilon).
\end{equation}

Helland \cite{helland1990partial}, N\ae{}s and Helland \cite{naes1993relevant}, and Naik and Tsai \cite{naik2000partial} gave a close form for the partial least squares (PLS) for the single-index model \eqref{e2},
\begin{eqnarray}\label{e3.9}
\beta_{\mathrm{PLS}}&=& R_q(R_{q}^T\Sigma_{\mathbf{x}}R_{q}\big)^-R_{q}^T\sigma_{xy}\\\label{e4}
& =&R_q(R_{q}^T\Sigma_{\mathbf{x}}R_{q}\big)^-R_{q}^T\Sigma_{\mathbf{x}}\beta_{\mathrm{OLS}},
\end{eqnarray}
where $\beta_{\mathrm{OLS}}$ is the coefficient estimator from ordinary least squares, $\Sigma_{\mathbf{x}}$ is the variance matrix of $\mathbf{x}$, $(\cdot)^-$ is the generalized inverse for singular matrices, and $R_{q}$ is the matrix of Krylov sequence,
\begin{equation}\label{e5}
R_{q}=(\sigma_{xy}, \Sigma_{\mathbf{x}}\sigma_{xy},\dots, \Sigma_{\mathbf{x}}^{q-1}\sigma_{xy}).
\end{equation}
$q$ is the number of columns if $R_{q}$ and will be discussed in the following section.

Followed the idea of single-index PLS, Li et al. \cite{li2007partial} proposed partial sliced inverse regression (PSIR) algorithm as an extension of SIR. The PSIR direction can be written as
\begin{equation}\label{e6}
\beta_{\mathrm{PSIR}}=P_{R_{q}^*(\Sigma_{\mathbf{x}})}\beta_{\mathrm{SIR}},
\end{equation}
where $\beta_{\mathrm{SIR}}$ is the SIR e.d.r.-direction, which is one-dimensional for the single-index model, and $P_{R_{q}^*(\Sigma_{\mathbf{x}})}$ is the projection matrix onto the Krylov subspace spanned by $R_{q}^*$ with respect to $\Sigma_{\mathbf{x}}$ inner product,
\begin{equation}\label{e7}
P_{R_{q}^*(\Sigma_{\mathbf{x}})}=R_{q}^*\big((R_{q}^*)^T\Sigma_{\mathbf{x}}R_{q}^*\big)^-(R_{q}^*)^T\Sigma_{\mathbf{x}}.
\end{equation}

Then, the PSIR direction can be further written as
\begin{eqnarray}\label{e9}
\beta_{\mathrm{PSIR}}&=&R_{q}^*\big((R_{q}^*)^T\Sigma_{\mathbf{x}}R_{q}^*\big)^-(R_{q}^*)^T\Sigma_{\mathbf{x}}\beta_{\mathrm{SIR}}\\\label{e9.1}
&=&R_{q}^*\big((R_{q}^*)^T\Sigma_{\mathbf{x}}R_{q}^*\big)^-(R_{q}^*)^T\omega,
\end{eqnarray}
where $\omega=\Sigma_{\mathbf{x}}\beta_{\mathrm{SIR}}$, similar to PLS, $R_{q}^*$ can be defined as
\begin{equation}\label{e8}
R_{q}^*=(\omega, \Sigma_{\mathbf{x}}\omega,\dots, \Sigma_{\mathbf{x}}^{q-1}\omega).
\end{equation}

The estimator of $\beta_{\mathrm{PSIR}}$, $\hat{\beta}_{\mathrm{PSIR}}$, can be calculated by replacing $\Sigma_{\mathbf{x}}$ and $\beta_{\mathrm{SIR}}$ in \eqref{e9} and \eqref{e8} by their sample estimator $\hat{\Sigma}_{\mathbf{x}}$ and $\hat{\beta}_{\mathrm{SIR}}$, respectively.


From \eqref{e4} and \eqref{e6}, one can see that PLS projects the coefficient from ordinary least squares onto the Krylov subspace with respect to $\Sigma_{\mathbf{x}}$ inner product, while PSIR projects the direction estimator from SIR, which consider the conditional expectation $E(\mathbf{x}|\mathbf{y})$. So PSIR should have better performance than PLS and SIR. It is shown that PSIR performs similar to or superior to PLS and much better than SIR when $n<p$ for regression, especially when the regression model is nonlinear or heteroscedastic \cite{li2007partial}.

\subsection{Parameter Selection}
For PSIR method, two parameters need to be determined. One is the number of slices $H$ in the sliced inverse regression, and the other is the $q$ for Krylov sequence \eqref{e8}.

First, it is suggested in \cite{li1991sliced} that the number of slices $H$ is not a crucial issue, since theoretical results showed that the SIR outputs do not change much for a wide range of $H$. Thus, $H$ was fixed at the most commonly used value $H = 10$ during the simulation.

There are several approaches estimating $q$ for either $R_q$ or $R_q^*$. N\ae{}s and Helland \cite{naes1993relevant} use the AIC as a criteria to select $q$. McQuarrie and Tsai \cite{McQTsai1998} and Naik and Tsai \cite{naik2000partial} use a corrected AIC to select $q$. A threshold approach by Li, et al. \cite{li2007partial} was used in this paper. Such approach is computationally simple and yields satisfactory results in the simulation.

Let
\begin{equation}
R_p=(\omega, \Sigma_{\mathbf{x}}\omega,\dots, \Sigma_{\mathbf{x}}^{p-1}\omega).
\end{equation}
and its estimator $\hat{R}_p$. Parameter $q$ can be estimated by the following formula.
\begin{eqnarray}
q & = & \sum_{j=1}^{p-1} I(r_j>\alpha), \\\nonumber
r_j & = & \lambda_j/\lambda_{j+1},\quad j=1,\dots,p-1,
\end{eqnarray}
where $I(\cdot)$ is the indicator function, $\lambda_j\ (j=1,\dots,p-1)$ are ordered eigenvalues of $\hat{R}_p\hat{R}_p^T$, and $\alpha$ is a prespecified threshold. In the simulation, the suggested value $\alpha=1.5$ \cite{li2007partial} was used.

\subsection{PSIR for Multiple-Index Model}
Only single-index model is discussed in the previous sections, but single-index partial sliced inverse regression can also be applied to the multiple-index model via space decomposition technique. 

To start with, for a given data $(\mathbf{y},\mathbf{x})$, perform the single-index PSIR to find the first direction $\beta_1$. The model can be written as $\mathbf{y}=g(\beta_1^T\mathbf{x},\mathbf{e_1})$, where $\mathbf{e_1}=(\mathbf{I}-\beta_1\beta_1^T)\mathbf{x}$, denoting the unexplained variation. Then, use $\mathbf{e_1}$ as the new predictors, and apply the single-index PSIR again to find the direction $\beta_2$ for the model $\mathbf{y}=g(\beta_2^T\mathbf{e_1},\mathbf{e_2})$, where $\mathbf{e_2}=(\mathbf{I}-\beta_2\beta_2^T)\mathbf{e_1}$. Keep doing such process until the unexplained variation is small enough. Therefore, the final e.d.r.-directions of $\mathbf{x}$ is $\big(\beta_1, \beta_2(\mathbf{I}-\beta_1\beta_1^T),\dots\big)$.

\section{X-space Decomposition}

\subsection{X-Space Decomposition of PLS}

PLS projects $(\mathbf{X}, \mathbf{Y})$ to the subspace spanned by the latent variables $(\mathbf{t}_1, \ldots, \mathbf{t}_A)$, where A is the number of PLS components:
\begin{equation} \label{e9.0}
  \left\{ \begin{array}{cl}
    \mathbf{X}= \mathbf{TP}^T + \mathbf{E}, \\
    \mathbf{Y}= \mathbf{TQ}^T + \mathbf{F}.
  \end{array} \right.
\end{equation}

The PLS scores and loadings can be calculated using nonlinear iterative partial least squares (NIPALS) algorithm \cite{Wold1975}, which determines the score vectors $\mathbf{t}$ and $\mathbf{u}$ iteratively. In each iteration, the loadings $\mathbf{p}$ and $\mathbf{q}$ are computed as
\begin{align*}
  \mathbf{p} &= \mathbf{X}^T\mathbf{t}/(\mathbf{t}^T\mathbf{t}), \\
  \mathbf{q} &= \mathbf{Y}^T\mathbf{u}/(\mathbf{u}^T\mathbf{u}).
\end{align*}
Matrix $\mathbf{X}$ is deflated before starting a new iteration:
\begin{equation}
  \mathbf{X} = \mathbf{X} - \mathbf{tp}^T.
\end{equation}
Note that there exists several PLS schemes and deflation of $\mathbf{Y}$ is not always needed \cite{Rosipal:Kramer:2006}. The relationship between $\mathbf{T}$ and original $\mathbf{X}$ is
\begin{equation*}
   \mathbf{T} = \mathbf{XR},
\end{equation*}
where $\mathbf{R} = \mathbf{W}(\mathbf{P}^T\mathbf{W})^{-1}$ along with $\mathbf{P}$ forms an oblique projector. The PLS decomposition of input variable space can be written as (\cite{li2010geometric})
\begin{eqnarray}
  \mathbf{x} &=& \hat{\mathbf{x}} + \tilde{\mathbf{x}}, \\\nonumber
  \hat{\mathbf{x}} &=& \mathbf{PR}^T \mathbf{x}, \\\nonumber
  \tilde{\mathbf{x}} &=& (\mathbf{I}-\mathbf{PR}^T)\mathbf{x}.
\end{eqnarray}

\subsection{X-space Decomposition of PSIR}

Similar to PLS, one can construct the space decomposition of PSIR. Let $\mathbf{P}_{\mathrm{PSIR}}$ be the loading matrix of PSIR. Given a new input vector $\mathbf{x}$, it can be decompose into
\begin{equation}\label{e12}
\mathbf{x}=\mathbf{P}_{s}\mathbf{x}+(\mathbf{I}-\mathbf{P}_{s})\mathbf{x}.
\end{equation}
where $\mathbf{P}_s$ is the projector onto the space spanned by $\mathbf{P}_{\mathrm{PSIR}}$,
\begin{equation}\label{e13}
\mathbf{P}_{s}=\mathbf{P}_{\mathrm{PSIR}}\big(\mathbf{P}_{\mathrm{PSIR}}^T\mathbf{P}_{\mathrm{PSIR}}\big)^{-1}\mathbf{P}_{\mathrm{PSIR}}^T.
\end{equation}

The dimension of $\mathbf{P}_{\mathrm{PSIR}}$ is $p \times r$, where $p$ is the number of process variables, and $r$ is the number of PSIR components. Usually, $r$ is relatively small and all the components in $\mathbf{P}_{\mathrm{PSIR}}$ are not highly correlated. Thus, it is possible to take inverse of $\mathbf{P}_{\mathrm{PSIR}}^T\mathbf{P}_{\mathrm{PSIR}}$. If the input data is highly correlated, one can use generalized inverse $(\cdot)^{-}$ instead of the regular inverse in \eqref{e13}.

In the single-index model \eqref{e2}, $r=1$ because only one direction will be chosen. $\beta_{\mathrm{PSIR}}$ in \eqref{e9} is used to denote the direction chosen by PSIR method. Moreover, the loading matrix $\mathbf{P}_{\mathrm{PSIR}}$ is degenerate to the loading vector $\beta_{\mathrm{PSIR}}$. Let
\begin{eqnarray}
\mathbf{t} & = & \big((\beta_{\mathrm{PSIR}}^T\beta_{\mathrm{PSIR}})^{-1}\beta_{\mathrm{PSIR}}^T\mathbf{x}\big)^T,\label{e16}\\
\mathbf{e} & = & \big(\mathbf{I}-\beta_{\mathrm{PSIR}}(\beta_{\mathrm{PSIR}}^T\beta_{\mathrm{PSIR}})^{-1}\beta_{\mathrm{PSIR}}^T\big)\mathbf{x}.\label{e17}
\end{eqnarray}
The decomposition of new vector $\mathbf{x}$ \eqref{e12} can be written as
\begin{eqnarray}\label{e14}
\mathbf{x} &=& \hat{\mathbf{x}} + \tilde{\mathbf{x}}, \\\nonumber
\hat{\mathbf{x}} &=&\beta_{\mathrm{PSIR}} \mathbf{t}^T,\\\nonumber
\tilde{\mathbf{x}} & = & \mathbf{e}.
\end{eqnarray}

\section{Process Monitoring based on PSIR}
Qin \cite{qin2003statistical} summarizes several statistical fault detection indices. Among them, two statistics, Hotelling's $T^2$ and the squared prediction error (SPE) are most commonly used to determine if a process is under normal condition.

\subsection{Hotelling's $T^2$}
Hotelling's $T^2$ is a generalized version of Student's $t$ statistic in the multivariate normal case. In the statistical process monitoring, $T^2$ is defined as following,
\begin{equation}\label{e15}
T^2=\mathbf{t}\Lambda^{-1}\mathbf{t}^T.
\end{equation}
where $\mathbf{t}$ is defined in \eqref{e16}. $\Lambda^{-1}$ is the sample covariance matrix of the score $\mathbf{T}$, $\Lambda^{-1}=\mathbf{T}^T\mathbf{T}/(n-1)$. If the input vector $\mathbf{x}$ has a multivariate normal distribution, the linear combination of its components $\mathbf{t}$ also has a multivariate normal distribution.

Based on the normal assumption, the Hotelling's $T^2$ follows a $F$ distribution,
\begin{equation}\label{e18}
\frac{n(n-r)}{r(n^2-1)}T^2 \sim F_{r,n-r},
\end{equation}
whose degrees of freedom is $r$, $n-r$. 

%

The $T^2$ statistic with its corresponding $F$ distribution can be used to test whether the score of a new input vector $\mathbf{x}$ has zero mean or not. If such test is failed for a given significance level, it is thought that there is a fault occurred in the score space.

In fact, in the single-index model, $\mathbf{t}$ degenerates to a univariate normal distribution. In addition, since the first degrees of freedom of $F$ is $r=1$, the $F$ distribution degenerates to the Student's $t$ distribution. For a given significance level $\alpha$, the upper control limit for $T^2$ is
\begin{equation}\label{e19}
\tau^2_{\alpha}=\frac{n^2-1}{n(n-1)}t_{n-1,\alpha}.
\end{equation}

The process is considered to be normal if $T^2 \leqslant \tau^2_{\alpha}$.

\subsection{Squared Prediction Error, SPE}
The squared prediction error is defined as the squared norm of the residual $\tilde{\mathbf{x}}$.
\begin{eqnarray}\nonumber
\mathrm{SPE} & = & \|\tilde{\mathbf{x}}\|^2\\\label{e21}
& = & \Big\|\big(\mathbf{I}-\beta_{\mathrm{PSIR}}(\beta_{\mathrm{PSIR}}^T\beta_{\mathrm{PSIR}})^{-1}\beta_{\mathrm{PSIR}}^T\big)\mathbf{x}\Big\|^2.
\end{eqnarray}

It can be treated as the squared distance of $\mathbf{x}$ to the space spanned by $\beta_{\mathrm{PSIR}}$.

Similar to Hotelling's $T^2$, assuming $\mathbf{x}$ has a multivariate normal distribution with zero mean and variance $\Sigma_{\mathbf{x}}$, thus, the residual $\tilde{\mathbf{x}}$ also has a multivariate normal distribution with zero mean and variance
\begin{equation}\label{e20}
\Sigma_{\mathrm{SPE}}=(\mathbf{I}-\mathbf{P}_{s})\Sigma_{\mathbf{x}}(\mathbf{I}-\mathbf{P}_{s})^T.
\end{equation}

Hence, $\mathrm{SPE}$ is a quadratic form of rank $p$. Box \cite{box1954some} gave an approximate distribution for the quadratic form \eqref{e21},
\begin{equation}
\mathrm{SPE} \sim g\chi^2(h),
\end{equation}
where $g$ is the weight and $h$ is the degrees of freedom,
\begin{eqnarray}\nonumber
g=\frac{\theta_2}{\theta_1}, & & h=\frac{\theta_1^2}{\theta_2},
\end{eqnarray}
and
\begin{eqnarray}\label{e21.5}
\theta_1=\sum_{i=1}^{p} \lambda_i, & & \theta_2=\sum_{i=1}^{p} \lambda^2_i,
\end{eqnarray}
where $\lambda_i$ are the eigenvalues of the variance matrix $\Sigma_{\mathrm{SPE}}$.


The SPE statistic can be used to test whether the residuals have zero means or not. For a given significance level $\alpha$, the upper control limit for SPE is
\begin{equation}\label{e22.1}
\delta_{\alpha}=g\chi^2_{\alpha}(h).
\end{equation}

If the SPE statistic for a new input vector $\mathbf{x}$ exceeds the upper limit $\delta_{\alpha}$, there is significance evidence shows the mean of the residual of such vector is not zero, which means the new input vector $\mathbf{x}$ lies outside of the space spanned by $\beta_{\mathrm{PSIR}}$, there is a fault occurred in the residual space.

Jackson and Mudholkar \cite{jackson1979control} gave another estimator for the upper limit of the quadratic form \eqref{e21} via normal approximation.
\begin{equation}\label{e23.1}
\delta^*_{\alpha}=\theta_1\bigg(1+\frac{\theta_2h_0(h_0-1)}{\theta_1^2}+z_{\alpha}\frac{(2\theta_2h_0^2)^{1/2}}{\theta_1}\bigg)^{1/h_0},
\end{equation}
where
\begin{eqnarray}\nonumber
h_0=1-\frac{2\theta_1\theta_3}{3\theta_2^2}, & & \theta_3=\sum_{i=1}^{p} \lambda_i^3,
\end{eqnarray}
and $z_{\alpha}$ is the upper $1-\alpha$ quantile of the standard normal distribution.

Jackson and Mudholkar's approximation \eqref{e23.1} is close to Box's \eqref{e22.1} when $\theta_2^2\approx \theta_1\theta_3$ (Nomikos and MacGregor \cite{nomikos1995multivariate}), this is not common, it may happen when there is a large principal component in $\Sigma_{\mathrm{SPE}}$. Therefore, Box's approximated upper control limit \eqref{e22.1} is recommended. The process is considered to be normal if $\mathrm{SPE} \leqslant \delta^2_{\alpha}$.

\subsection{Combined Index $\varphi$}
$T^2$ captures the faults occurred in the score space whereas SPE captures the faults occurred in the residual space. Therefore, a combined index proposed by Yue and Qin \cite{yue2001reconstruction} is used in this paper.

The combined index $\varphi$ combines the $T^2$ and SPE statistics, it is defined as follows,
\begin{equation}\label{e26}
\varphi=\frac{T^2}{\tau_{\alpha}^2}+\frac{\mathrm{SPE}}{\delta_{\alpha}^2}.
\end{equation}

Yue and Qin \cite{yue2001reconstruction} gave a control limit $\zeta^2$ for the combined index $\varphi$ \eqref{e26}.
\begin{equation}\label{e27}
 \zeta^2_{\alpha}=g_{\varphi}\chi_{\alpha}^2(h_{\varphi}),
\end{equation}
with confidence level $(1-\alpha)\times100$, $g_{\varphi}$ and $h_{\varphi}$ are defined as
\begin{eqnarray}
 g_{\varphi} &=& \bigg(\frac{1}{\tau^4}+\frac{\theta_2}{\delta^4}\bigg)\bigg/\bigg(\frac{1}{\tau^2}+\frac{\theta_1}{\delta^2}\bigg),\\
 h_{\varphi} &=& \bigg(\frac{1}{\tau^2}+\frac{\theta_1}{\delta^2}\bigg)^2\bigg/\bigg(\frac{1}{\tau^4}+\frac{\theta_2}{\delta^4}\bigg),
\end{eqnarray}
$\theta_1$ and $\theta_2$ are defined in \eqref{e21.5}.

The process is considered to be normal if $\varphi \leqslant \zeta^2_{\alpha}$.

\section{Simulation Studies}

The purpose of the simulation is to compare the fault detection rates among PLS, SIR, and PSIR by Monte Carlo simulation. Two single-index models were considered, one was linear and the other was nonlinear,
\begin{eqnarray}\label{e23}
y_1 & = & x_1+x_2+\dots+x_{10}+\sigma_1\epsilon,\\\label{e24}
y_2 & = & \frac{\exp(x_1+x_2+\dots+x_{10})}{1+\exp(x_1+x_2+\dots+x_{10})}+\sigma_2\epsilon,
\end{eqnarray}
where $\mathbf{x}=(x_1, x_2, \dots, x_{10})^T$ followed a multivariate normal distribution with zero means, the variance of each $x_i$ was assumed to be one, and the covariance of each pair $(x_i, x_j), i\neq j$ was 0.5, i.e, $\mathbf{x}\sim\mathcal{N}_{10}(\mathbf{0},\Sigma)$, 
\begin{equation*}
\Sigma=\left(\begin{array}{ccccc}
              1 \\
	       & 1 & & \text{\large 0.5}\\
	       & & \ddots\\
	       & \text{\large 0.5} & & 1\\
	       & & & & 1\\
             \end{array}
\right).
\end{equation*}

$\epsilon$ in \eqref{e23} and \eqref{e24} were random noises, which were assumed to have standard normal distributions and were independent with $\mathbf{x}$, $\sigma_1$ and $\sigma_2$ were the standard deviation of the random noises in order to make a reasonable noise magnitude. In the simulation. $\sigma_1$ and $\sigma_2$ were chosen to be $1/20$ of the standard deviation of $y_1$ and $y_2$, respectively.

The simulated faulty samples $\mathbf{x^*}$ were in the form 
\begin{equation}\label{e25}
\mathbf{x^*}=\mathbf{x}+f\times \mathbf{\xi},
\end{equation}
where $\mathbf{\xi}$ was the fault direction to be added, and $f$ was the fault magnitude.

To perform the simulation, 100 fault directions with unit magnitude were generated first. Then, 10 Monte Carlo simulations were run for each fault direction. Within each simulation, 500 regular samples were generated, followed by another 100 faulty samples produced by \eqref{e25}. Three method, PLS, SIR, and PSIR, were performed to detect the fault occurrence. The mean of the fault detection rates based on the combined index \eqref{e26} and its control limit \eqref{e27} as well as the standard deviation of the detection rates were calculated.

\begin{table}[p!]
\caption{Means of the Fault Detection Rates (\%)}
\begin{center}
\begin{tabular}{lllllll}
\hline
 & \multicolumn{3}{c} {Linear Model} &  \multicolumn{3}{c} {Nonlinear Model}\\
\cline{2-7}
$f$  & PLS & SIR & PSIR & PLS & SIR & PSIR \\
\hline
0 & 1.41 & 1.51 & 1.51 & 1.41 & 1.86 & 1.80 \\ 
1 & 2.16 & 2.26 & 3.17 & 2.16 & 2.75 & 3.84 \\ 
2 & 5.51 & 5.71 & 7.03 & 5.51 & 6.50 & 8.10 \\ 
3 & 14.35 & 14.57 & 16.36 & 12.54 & 15.83 & 15.99 \\ 
5 & 54.08 & 55.00 & 55.26 & 51.09 & 55.21 & 57.72 \\ 
8 & 97.03 & 97.02 & 98.86 & 97.05 & 96.93 & 99.03 \\ 
12 & 100 & 100 & 100 & 100 & 100 & 100 \\ 
\hline
\end{tabular}
\end{center}
\label{t1}
\vspace{2em}
\caption{Standard Deviations of the Fault Detection Rates (\%)}
\begin{center}
\begin{tabular}{lllllll}
\hline
 & \multicolumn{3}{c} {Linear Model} &  \multicolumn{3}{c} {Nonlinear Model}\\
 \cline{2-7}
 $f$  & PLS & SIR & PSIR & PLS & SIR & PSIR \\
\hline
0 & 1.24 & 1.26 & 1.54 & 1.24 & 1.41 & 1.83  \\
1 & 1.46 & 1.49 & 1.79 & 1.46 & 1.68 & 2.11 \\ 
2 & 2.39 & 2.47 & 2.80 & 2.39 & 2.65 & 3.15 \\ 
3 & 4.12 & 4.15 & 4.43 & 4.10 & 4.45 & 4.47 \\ 
5 & 10.05 & 10.02 & 8.95 & 10.05 & 9.78 & 8.19 \\ 
8 & 3.55 & 3.48 & 1.85 & 3.54 & 3.40 & 0.86 \\ 
12 & 0 & 0 & 0 & 0 & 0 & 0 \\ 
\hline
\end{tabular}
\label{t2}
\end{center}
\end{table}

Table \ref{t1} presents the means of the fault detection rates for PLS, SIR, and PSIR methods in both the linear model \eqref{e23} and nonlinear model \eqref{e24}, with the fault magnitudes from 0 to 15. Table \ref{t2} presents the corresponding standard deviations from the simulation.

From Table \ref{t1}, it is observed that for all the three methods, the fault detection rates increased when the fault magnitude became larger, and the fault detection rates eventually reached 100\% for a large fault magnitude. By comparing the fault detection rates in each fault magnitude, the inverse regression based method PSIR and SIR performed better than PLS, especially in nonlinear model, and PSIR had the best fault detection rates among these three methods.

\section{Conclusion and Discussion}

Sliced inverse regression is a popular dimension reduction method in computer science, biology, social science, and economics. In this paper, the methodology of the partial sliced inverse regression is used for quality-relevant multivariate process monitoring. Li, et al. \cite{li2007partial} showed PSIR can gain information on both conditional mean $E(\mathbf{y}|\mathbf{x})$ and conditional variance $\mathrm{var}(\mathbf{y}|\mathbf{x})$, while PLS can only retrieve information from conditional mean. The simulation study in Section 5 confirmed that PSIR had better fault detection rates in both linear and nonlinear model. 

Because the fault detection is only performed in the X-space, the superiority of the inverse regression based methods is limited. But in process monitoring, the responses should definitely be taken into account because they are quality variables which contains the information of processing conditions. 

Moreover, comparing with PLS, which is the most commonly used method in this area, PSIR is computationally simple since it requires no iterations for finding the loading matrix. Therefore, PSIR is the most advantage method for multivariate statistical process monitoring.

\bibliographystyle{plain}
\bibliography{psir}

\begin{thebibliography}{10}

\bibitem{box1954some}
G.E.P. Box.
\newblock {Some theorems on quadratic forms applied in the study of analysis of
  variance problems, I. Effect of inequality of variance in the one-way
  classification}.
\newblock {\em The Annals of Mathematical Statistics}, 25(2):290--302, 1954.

\bibitem{cook1991discussion}
R.D. Cook and S.~Weisberg.
\newblock {Discussion of Li (1991)}.
\newblock {\em Journal of the American Statistical Association}, 86:328--332,
  1991.

\bibitem{duan1991slicing}
N.~Duan and K.C. Li.
\newblock {Slicing regression: a link-free regression method}.
\newblock {\em The Annals of Statistics}, 19(2):505--530, 1991.

\bibitem{dunia1998subspace}
R.~Dunia and S.~Joe~Qin.
\newblock {Subspace approach to multidimensional fault identification and
  reconstruction}.
\newblock {\em AIChE Journal}, 44(8):1813--1831, 1998.

\bibitem{eaton1986characterization}
M.L. Eaton.
\newblock {A characterization of spherical distributions}.
\newblock {\em Journal of Multivariate Analysis}, 20(2):272--276, 1986.

\bibitem{helland1990partial}
I.S. Helland.
\newblock {Partial least squares regression and statistical models}.
\newblock {\em Scandinavian Journal of Statistics}, 17(2):97--114, 1990.

\bibitem{jackson1979control}
J.E. Jackson and G.S. Mudholkar.
\newblock {Control procedures for residuals associated with principal component
  analysis}.
\newblock {\em Technometrics}, 21(3):341--349, 1979.

\bibitem{kresta1991multivariate}
J.V. Kresta, J.F. Macgregor, and T.E. Marlin.
\newblock {Multivariate statistical monitoring of process operating
  performance}.
\newblock {\em The Canadian Journal of Chemical Engineering}, 69(1):35--47,
  1991.

\bibitem{li2010geometric}
G.~Li, S.J. Qin, and D.~Zhou.
\newblock {Geometric properties of partial least squares for process
  monitoring}.
\newblock {\em Automatica}, 46(1):204--210, 2010.

\bibitem{li1991sliced}
K.C. Li.
\newblock {Sliced inverse regression for dimension reduction}.
\newblock {\em Journal of the American Statistical Association}, pages
  316--327, 1991.

\bibitem{li1991rejoinder}
K.C. Li.
\newblock Sliced inverse regression for dimension reduction: Rejoinder.
\newblock {\em Journal of the American Statistical Association},
  86(414):337--342, 1991.

\bibitem{li2003dimension}
K.C. Li, Y.~Aragon, K.~Shedden, and C.~Thomas~Agnan.
\newblock {Dimension reduction for multivariate response data}.
\newblock {\em Journal of the American Statistical Association},
  98(461):99--109, 2003.

\bibitem{li2007partial}
L.~Li, R.D. Cook, and C.L. Tsai.
\newblock {Partial inverse regression}.
\newblock {\em Biometrika}, 2007.

\bibitem{LiYin2008}
Lexin Li and Xiangrong Yin.
\newblock Sliced inverse regression with regularizations.
\newblock {\em Biometrics}, 64(1):124--131, 2008.

\bibitem{McQTsai1998}
Allan D.~R. McQuarrie and Chih-Ling Tsai.
\newblock {\em Regression and Time Series Model Selection}.
\newblock World Scientific Publishing Company, 1998.

\bibitem{naes1993relevant}
T.~N\ae{}s and I.S. Helland.
\newblock {Relevant components in regression}.
\newblock {\em Scandinavian Journal of Statistics}, 20(3):239--250, 1993.

\bibitem{naik2000partial}
P.~Naik and C.L. Tsai.
\newblock {Partial least squares estimator for single-index models}.
\newblock {\em Journal of the Royal Statistical Society: Series B (Statistical
  Methodology)}, 62(4):763--771, 2000.

\bibitem{nomikos1995multivariate}
P.~Nomikos and J.F. MacGregor.
\newblock {Multivariate SPC charts for monitoring batch processes}.
\newblock {\em Technometrics}, 37(1):41--59, 1995.

\bibitem{qin2003statistical}
S.J. Qin.
\newblock {Statistical process monitoring: basics and beyond}.
\newblock {\em Journal of Chemometrics}, 17(8-9):480--502, 2003.

\bibitem{Rosipal:Kramer:2006}
Roman Rosipal and Nicole Kr\"amer.
\newblock Overview and recent advances in partial least squares.
\newblock In Craig Saunders, Marko Grobelnik, Steve Gunn, and John
  Shawe-Taylor, editors, {\em Subspace, Latent Structure and Feature
  Selection}, volume 3940 of {\em Lecture Notes in Computer Science}, pages
  34--51. Springer Berlin / Heidelberg, 2006.

\bibitem{Wold1975}
H.~Wold.
\newblock {\em Path Models with latent variables: The NIPALS approach.}
\newblock Acad. Pr., New York, NY, 1975.

\bibitem{ye2010sliced}
Z.~Ye and J.~Yang.
\newblock {Sliced inverse moment regression using weighted chi-squared tests
  for dimension reduction}.
\newblock {\em Journal of Statistical Planning and Inference},
  140(11):3121--3131, 2010.

\bibitem{yue2001reconstruction}
H.H. Yue and S.J. Qin.
\newblock {Reconstruction-based fault identification using a combined index}.
\newblock {\em Industrial \& engineering chemistry research},
  40(20):4403--4414, 2001.

\bibitem{Zhong2005RSIR}
Wenxuan Zhong, Peng Zeng, Ping Ma, Jun~S. Liu, and Yu~Zhu.
\newblock Rsir: regularized sliced inverse regression for motif discovery.
\newblock {\em Bioinformatics}, 21(22):4169--4175.

\end{thebibliography}

\end{document}